\documentclass{article}

\usepackage{arxiv}
\usepackage{hyperref}       
\usepackage{url}            
\usepackage{booktabs}       
\usepackage{amsfonts}       
\usepackage{nicefrac}       
\usepackage{microtype}      
\usepackage{xcolor}      

\usepackage{graphicx}
\usepackage{amsmath}
\usepackage{amsthm}
\usepackage{booktabs}
\usepackage{algorithm}
\usepackage{algorithmic}

\usepackage{subfigure}  
\usepackage{comment}
\usepackage{bm}
\usepackage{multirow}

\title{Melody Structure Transfer Network: Generating Music with Separable Self-Attention}

\author{
Ning Zhang
\and
Junchi Yan$^*$
\affiliations
Department of Computer Science and Engineering, Shanghai Jiao Tong University
\emails
\{ningz, yanjunchi\}@sjtu.edu.cn
}

\begin{document}

\maketitle

\begin{abstract}
  Symbolic music generation has attracted increasing attention, while most methods focus on generating short piece (mostly less than 8 bars, and up to 32 bars). Generating long music calls for effective expression of the coherent music structure. Despite their success on long sequences, self-attention architectures still have challenge in dealing with long-term music as it requires additional care on the subtle music structure. In this paper, we propose to transfer the structure of training samples for new music generation, and develop a novel separable self-attention based model which enable the learning and transferring of the structure embedding. We show that our transfer model can generate music sequences (up to 100 bars) with interpretable structures, which bears similar structures and composition techniques with the template music from training set. Extensive experiments show its ability of generating music with target structure and well diversity. The generated 3,000 sets of music is uploaded as supplemental material.
\end{abstract}

\section{Introduction}
Algorithmic composition has a standing history since the early stochastic composition system `Atree'~\cite{xenakis1963formalized}. Handcrafted grammars and rules are used to regularize the generation of music of different styles and structures~\cite{mclean2018oxford,nierhaus2009algorithmic}. Developing these rules is nontrivial as the music theory itself contains lots of rules, and those factors contributing to the music art of different genres are difficult to specify. The rules vary with composers' style and music forms. \cite{ebciouglu1988expert} builds a four-part J.S. Bach chorale composition system according to over 350 handcrafted rules. \cite{cope2000algorithmic} extends it with learning from the score corpus of specific composer and creates its own grammar and rules. However, the algorithmic composition still remains open~\cite{briot2018deep}.

Recent deep music generation models~\cite{briot2018deep,briot2017deep} can generate quality short pieces. While for long sequence, the generated pieces suffer unclear structure. A complete music piece usually consists of more than one track of long note sequences, which has clear and complex horizontal structures (along the time axis) and vertical structures (between multiple instruments or tracks). The horizontal structures are related to the music form, which consists of three-level of structures: sub-phrase, phrase, and section. The sub-phrase contains the basic ideas (motives), which are then developed into phrases and sections according to some composition techniques~\cite{xenakis1963formalized}. 

The rhythm, theme and emotion of the music piece evolve over time with the extension of sub-phrase and phrases. The organization of music elements into music pieces consists of many rigorous rules. A small violation e.g. one more or less beat, can wreck the structure~\cite{mclean2018oxford}.

Deep music generation approaches can be categorized by the form of music representation. One category is the piano roll based models. The piano roll represents the symbolic music as images of shape $P\times T\times I$, where $P, T, I$ denote the number of pitches, time steps and instruments, respectively. Some works~\cite{huang2017counterpoint,dong2018musegan,dong2018convolutional} build models to generate piano rolls. Limited by the image characteristics, these methods generate music of short length (less than 16 bars). Methods falling into the second category treat music as event sequences, including note based sequences and frame based sequences. Then music generation can be viewed as  sequence generation, and various sequence models are used to describe the joint distribution of the music sequence. Many works also utilize RNN to model the music sequences~\cite{jaques2017tuning,sturm2016music}, and the length can be up to 512 tokens~\cite{simon2017performance}. Recently, the self-attention architectures have demonstrated their superiority on long-term sequence processing. Methods \cite{huang2018music,musenet,huang2020pop} apply the self-attention architectures (transformers)~\cite{vaswani2017attention} to music sequence generation. It has been shown in \cite{huang2018music} for generating music of 2048 tokens.

There exists structure coherence for long sequence music. Though \cite{huang2018music} claims that it could generate music with better structure than \cite{simon2017performance}, it only qualitatively shows a sample with more repeated patterns. Other works~\cite{medeot2018structurenet,roberts2018hierarchical} stress the problem of music structure in music generation. They try to enhance structure by increasing the probability that the model generates repeated elements. However, more repeat patterns do not mean better structure. Human compositions repeat and vary the motives to express some certain emotions.  \cite{lattner2018imposing} also proposes to transfer the structure from a template music to generate new music piece. However, it requires to design constraints and train a model for every template music. This limits its applicability.



In this paper, we present a model which can transfer the structures of template music pieces to generate new pieces. Different from \cite{lattner2018imposing} in which the model is designed with hand-crafted constraints to learn the structure of one specific music piece less than 32 bars, our model can automatically learn and transfer the structures of all pieces up to 100 bars in the training set. We make two important observations to motivate our approach:

1) Music structure transfer from real music to generated pieces can be label-free and (potentially) an efficient approach, which in fact has not been well-studied in literature;

2) By examining the transformer based music generation models, we find that the learned self-attention matrix closely relates to music structure. In another word, transferring self-attention relations can help transfer music structure. This motivates to explore the better use of self-attention mechanism.

\begin{figure}[tb!]
  \centering
  \includegraphics[width=1.0\linewidth]{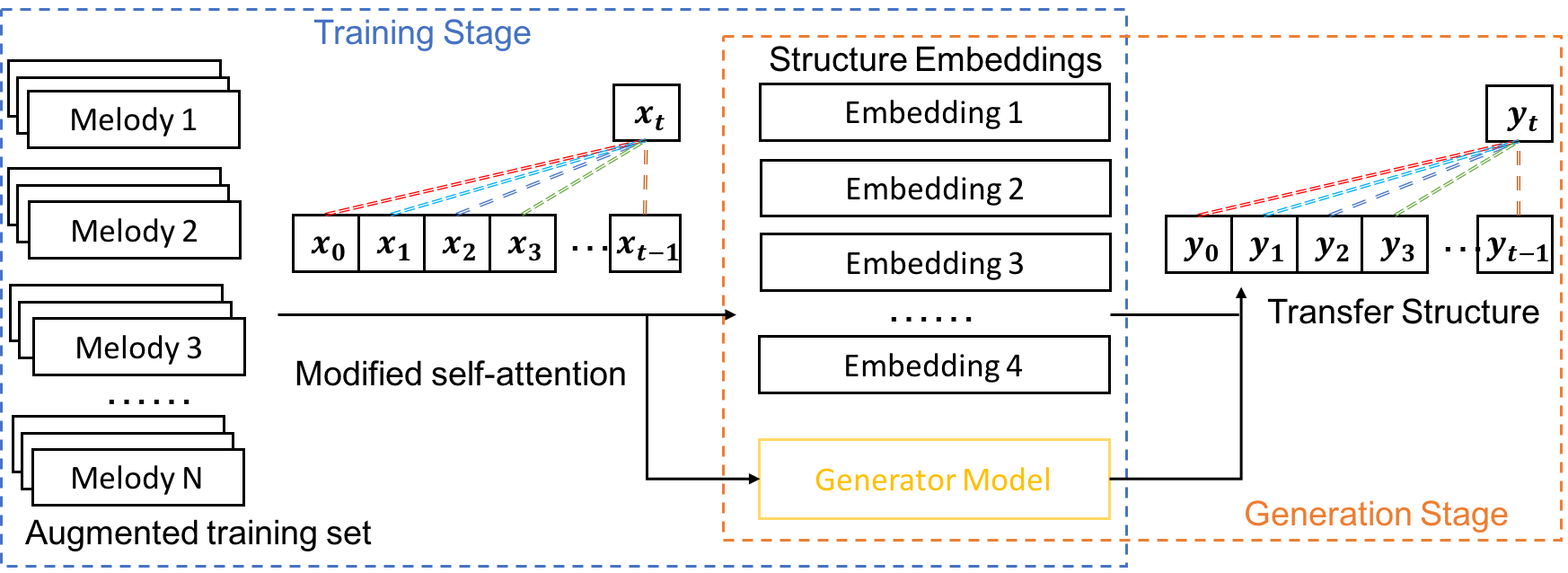}
  \vspace{-10pt}
  \caption{Overview of Melody Structure Transfer Network (MSTN).}
  \label{fig:overview}
\end{figure}

In this paper, we present novel separable self-attention mechanism based models to encode the structure information from the real musics into embeddings, and these embeddings can be transferred to the generation process to achieve realistic long music generation as shown in Fig.~\ref{fig:overview}. Besides, different from other transformer based models~\cite{huang2018music,musenet,huang2020pop} which take the note-based event sequence representations, we utilize the frame-based event sequence to represent the music scores, which simplifies the control of the metrical structures of music. Moreover, we introduce the key signature tokens to control the tonality of the generated music. Experiments show that our method can transfer the structure of templates to the generation of new pieces. The trained model can develop any given motives into a new piece by using similar composition techniques with the templates. Also, the model can generate diverse pieces, rather than simply remember the sequence of the training data. The contributions of this work are:

1) Differing from~\cite{huang2018music,medeot2018structurenet} which only try to increase the probability of repeated patterns, we propose to transfer structure of training samples for music generation especially for long sequence. Unlike \cite{lattner2018imposing}, our method could transfer the structures of all the training pieces and does not rely on hand-crafted constraints.

2) We develop a transferable self-attention mechanism to achieve effective structure transfer even for very long music. It separates the computation of query, key and value, to fulfill flexible structure transfer. Note that~\cite{huang2018music} expects the self-attention to learn all the structures from the training set. In contrast, our tailored self-attention supports the learning of structure embedding from each of the training data, and transfers it effectively to the generated samples.

3) We propose several quantitative metrics based on music theory to measure the rhythm structure similarity and interval structure similarity, as well as the diversity, for generated music. Extensive quantitative experiments show the effectiveness of our proposed structure transfer network in generating new music pieces (up to 100 bars). Besides, we quantitatively show that the generated pieces are of well diversity, rather than simply remember the training pieces.

\section{Preliminaries and Related Works}\label{sec:relatedworks}
\paragraph{Music Generation by Sequence Models}
Existing music generation models can be categorized into piano roll based models and event sequence based models. We briefly describe the latter ones, which are related to our work. In these models, music scores are represented as an event sequence $\mathbf{s}=[s_1, s_2,...,s_t,...,s_n]$, where $s_i$ is the event token at time step $i$. Its joint distribution can be factorized into: $    p_{\theta}(\mathbf{s})=p_{\theta}(s_1)p_{\theta}(s_2|s_1)...p_{\theta}(s_n|s_{n-1}...s_{1})$.

Various sequence models have been developed to describe the conditional probabilities. From the pioneering music generation system CONCERT~\cite{mozer1994neural}, to recently Celtic system \cite{sturm2016music} and PerformanceRNN \cite{oore2018time}, which employs RNN to generate symbolic music. Meanwhile, self-attention architecture based models e.g. music transformer~\cite{huang2018music} and MuseNet~\cite{musenet} are also devised to generate long sequence music.

These models are typically trained with the teacher-forcing strategy~\cite{bengio2015scheduled} to predict the next token. Sampling is performed from the conditional distribution step by step, to generate sequences of arbitrary length. However, the sequence models receive no extra input related to music structure or other information, resulting poor structures of the generated samples. Though the self-attention models often outperform LSTM models for generating long-term music~\cite{huang2018music}, whereby the motives can be repeated along the whole sequences, their generated results still suffer unclear structure. In conclusion, the problem for long-term music generation remains open and far from resolved.

\paragraph{Self-attention Mechanisms}
Transformer~\cite{vaswani2017attention} based models mostly directly employ stacked self-attention blocks (like GPT and GPT-2~\cite{radford2019language}) to generate music auto-regressively. 

The attention layer first calculates three matrices: queries ($Q$), keys ($K$) and values ($V$) from input embedding sequence $X^{i} = (x_0, x_1, x_2, ..., x_L)$, 
where $x_t\in \mathcal{R}^D$ for time step $t$, and $i$ is the index of transformer blocks and $L$ is the length of input sequence. $Q = XW_{Q}$,
$K=XW_{K}$ and $V=XW_{V}$. Here $W_{Q}$, $W_{K}$ and $W_{V}$ are $D\times D$ matrices. Then $Q$, $K$ and $V$ are $L\times D$ matrices, each of which is then split into $H\cdot L \times D_h$ attention heads, where $D_h = \frac{D}{H}$. The multi-heads mechanism allow the model to focus on different parts of the history. The self attention is computed as:
\begin{equation}\label{eq:attention}
    Z = Attention(Q, K, V) = Softmax\left(\frac{QK^{\top}}{\sqrt{D_h}}\right)V 
\end{equation}
The subsequent layers transform $Z$ to get each block's output $X^{i+1}$ (input to next block). Initially, $X^0 = E_x + E_{pos}$, where $E_x$ is the embedding sequence of input tokens and $E_{pos}$ is the positional encoding. An upper triangular mask ensures queries cannot attend to keys later in the sequence.

\section{The Proposed Approach}\label{sec:proposed}
\subsection{Connecting Self-attention and Structure}
The self-attention can be viewed as a generalization of self-similarity matrix (SSM)~\cite{huang2018music}. SSM can reflect the self-similarity structure of a music piece, and has been used for music segmentation and motif discovery~\cite{bello2011measuring,jun2015music}, and music generation~\cite{jhamtani2019modeling}. However, SSM can only capture the repetition patterns. According to the composition theory~\cite{xenakis1963formalized}, there are more than 10 kinds of techniques e.g. transposition, compression, expansion, mirror for developing motif into melody.

We further identify the connections between self-attention and the sequence structure. The multi-head self-attention mechanism allows modelling of multi-level dependencies from input sequences~\cite{vaswani2017attention}. Each position in the decoder attends to all positions in the decoder up to (including) that position. This is somewhat similar to the composition process by human, where a prime motif (a short sequence of several notes) is first devised and then developed into a long sequence via repetitions and variations. These repetitions and variations occur at multiple levels (sub-phrases, phrases and sections) and form the multi-level structures. The multi-level dependencies learned by the self-attention mechanism is closely related to the multi-level structures. 

Seeing the fundamental connection between self-attention and sequence structure, we argue that the learned dependencies in self-attention mechanism can be used to guide the generation of new music piece, and the realistic structures can be kept during generation by dependency transfer.

\subsection{Dependency Transfer via Attention}
Now the problem becomes how to transfer the learned dependencies from the real music to generation. Specifically, the generation of token $s_t$ in step $t$ should attend to similar positions as the real one. To achieve this goal, we hope the calculation of self-attention can be expressed as:
\begin{equation}\label{eq:newatt}
\begin{array}{cl}
    Z = Attention (Q, K, V) &=Att_{M} \cdot V\\ &=\mathbf{f}(dep, \mathbf{g}(s)) \cdot V
\end{array}
\end{equation}
where $Z$ is the output hidden state of a self-attention layer, and $Att_{M}$ is the attention matrix at this position. $s$ denotes the past tokens in the sequence, $dep$ is the structure dependency related variables for transfer. $\mathbf{f}(\cdot)$, $\mathbf{g}(\cdot)$ are transformations. 

We further expect that $dep$ shall meet two requirements: 1) containing rich structure-related information; 2) independent from the past input sequences thus it can be transferred to different generated sequences.

For the first requirement, as the structure is closely related to positions, to force variable $dep$ to learn the structure related information, we make only  $dep$ relate to the positions. Both $V$ and $\mathbf{g}(s)$ should be independent from the positions. To satisfy the second requirement, the computation of $dep$ should not include the input tokens.

Note that directly transplanting the self-attention matrix $Att_{M}$ from a training piece to the generation phrase can hardly work well. Because in the original self-attention mechanism, the calculation of attention in each layer depends on the input token at each time step (think back the calculation of query, key and value). Different input sequences will lead to different attention matrices. In this paper, we show how to implement the self-attention mechanism in Eq.~\ref{eq:newatt} by separating the calculation of query, key and value. A structure embedding is introduced to help the learning of variable $dep$. 

\subsection{Design of Transferable Self-attention}
\begin{figure}[tb!]
  \centering
  \includegraphics[width=1.0\linewidth]{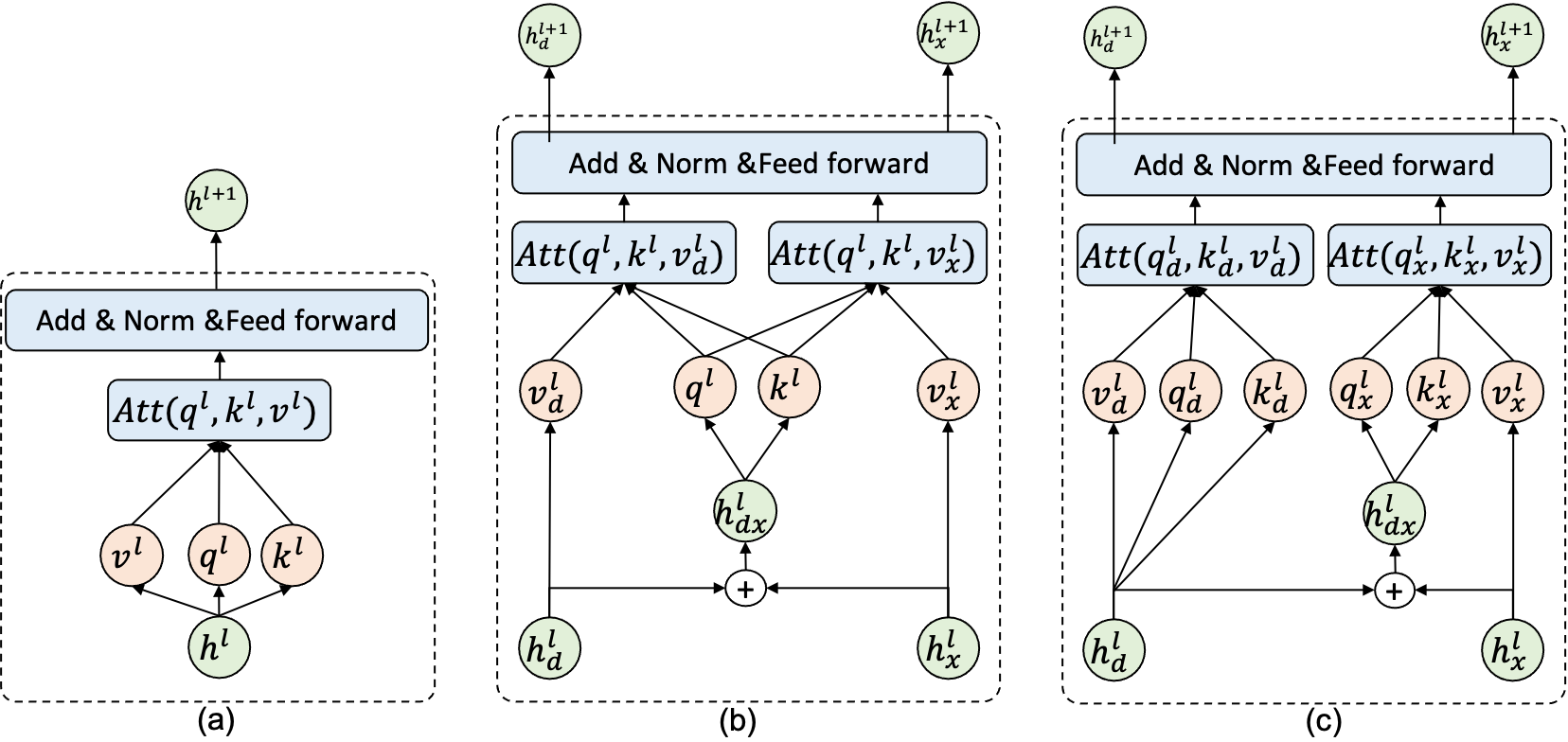}
  \vspace{-10pt}
  \caption{(a) The original self-attention block and (b, c) the proposed separable self-attention block. $h^l_{d}$ denotes the structure hidden state of the $l^{th}$ block and $h^l_{d}$ denotes the note hidden state of this block. The computation process of (b) and (c) correspond to Eq.~\ref{eq: newatt1} and \ref{eq: newatt2} (The superscript $l$ is omitted for clarity).
}
  \label{fig:newatt3}
\end{figure}

In the original self-attention mechanism~\cite{vaswani2017attention}, the computations of query, key, value all depend on the position embedding and input token embedding. Here, we separate the calculation of query, key, value, and introduce a structure hidden state $h_d$ and note hidden state $h_x$. We design two architectures to implement the separable self-attention mechanism in Eq.~\ref{eq:newatt}. The designed architectures are shown in Fig.~\ref{fig:newatt3}. The calculation of the separable self-attention are:
\begin{equation}\label{eq: newatt1}
\begin{array}{ll}
  
     q=(h_d+\lambda h_x)W_q & k=(h_d+\lambda h_x)W_k  \\
     v_d=h_d W_d  & v_x =h_x W_x\\
     a_d=Att(q,k)v_d & a_x = Att(q,k)v_x
\end{array}
\end{equation}
\begin{equation}\label{eq: newatt2}
\begin{array}{ll}
     q_x=(h_d+\lambda h_x)W_{q_x} & k_x=(h_d+\lambda h_x)W_{k_x}   \\
     q_d=h_d W_{q_d} & k_d = h_d W_{k_d}  \\
     v_d=h_d W_{v_d}  & v_x = h_x W_{v_x}\\
     a_d=Att(q_d,k_d)v_d & a_x = Att(q_x,k_x)v_x
\end{array}
\end{equation}
where $ Att(q, k) = Softmax\left(\frac{qk}{\sqrt{D_h}}\right)$.

Note Eq.~\ref{eq: newatt1} and \ref{eq: newatt2} specify the separable self-attention mechanism. They corresponds to the architectures in Fig.~\ref{fig:newatt3}(b) and (c), respectively (We omit the superscript $l$ for clarity). Different from the original self-attention block where the only hidden state $h^l$ encode all the information from input, we introduce separated structure hidden states $h_d$ and the note hidden states $h_x$. The $v_x$ and $v_d$ at each block are calculated from the input note hidden states and structure states, respectively. In Fig.~\ref{fig:newatt3}(b) the attention coefficients for $v_x$ and $v_d$ are the same, the $q$ and $k$ could be related to both the note states and structure states. In Fig.~\ref{fig:newatt3}(c) the attention coefficients for $v_x$ and $v_d$ are different, $q_d$ and $k_d$ are only related to the structure states, while $q_x$ and $k_x$ still could be related to both the note states and structure states. 

At the first block, the original music transformer architecture~\cite{radford2019language} takes the addition of note embedding $E_{note}$ and position embedding $E_{pos}$ as input, where $h^0 = E_{note} + E_{pos}$. We introduce a structure embedding $E_{d}$: 
\begin{equation}\label{eq:hid}
   h^0_{d} = E_{pos} + E_{d}, \quad h^0_{x} = E_{note} .
\end{equation}

Fig.~\ref{fig:newatt3}(b) and (c) are two exemplar implementations of Eq.~\ref{eq:newatt}. Others may design more architectures for Eq.~\ref{eq:newatt}.

\subsection{Structure Embedding for Transfer}

The structure embedding in Eq~\ref{eq:hid} is defined as $E_d=\{e_{d_t}|t=0,1,2,...,T\}$, where $e_{d_{t}}$ is a trainable vector which denotes the embedding at $t^{th}$ time step for a training piece. It is computed as 
\begin{equation}
    e_{d_{t}} = w_{t}\cdot e_{d} 
\end{equation}
where $w_{t}$ is the transform vector of size $(n_{state},)$ at $t^{th}$ time step, $e_{d}$ is the structure representation of a training piece. Each music piece in the original training set is assigned a structure representation of size $(n_{state},)$ randomly. $n_{state}$ is the length of hidden state in the transformer architecture.

We augment the music pieces in original training set using the pitch transposition technique~\cite{huang2018music}, which is to transpose the pitches totally by $\{-6, -5,...,5,6\}$ steps. This augmentation changes the tonality of the original music, but not the structure. Thus the  augmented versions share the same structure embedding $e_{d}$ with the original ones. The $e_{d}$ for each training piece is learned during training. Then the learned $e_{d}$ can be transferred to the generation stage.

\subsection{Advantages to Conditional Transformer}
Someone may argue that a conditional music transformer (CMT)~\cite{keskar2019ctrl} may also work for the structure transfer. In this setting, the structure embedding is used as a control code, which is prepended ahead of the music sequence embeddings. This code provides a control over the generation process. However, the control embedding can only affect the global dependencies via its keys at that single position. Therefore, it has difficulty in controlling the subtle dependencies between all the tokens along the sequence and thus lead to unstable structure transfer. We also implement a conditional transformer, as compared in our experiments.

\subsection{Data Representation for Music Generation}\label{sec:representation}
We adopt the frame-based event representation scheme in~\cite{pati2019inpainting}: time is quantized using uneven sub-division. Unlike other works quantizing uniformly with the sixteenth note, the work \cite{pati2019inpainting} designs the uneven subdivision scheme, where each beat is divided into 6 uneven ticks. This scheme allows to represent note sequences including triplets efficiently. In addition, real note names are used for generation of readable sheet music.

\begin{figure}[tb]
  \centering
  \includegraphics[width=1.0\linewidth]{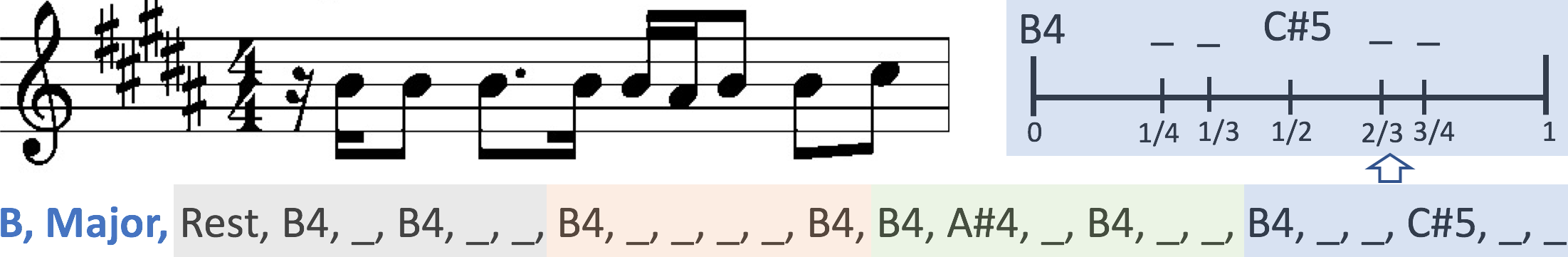}
  \caption{Representation of a sheet music segment (reproduced from \protect\cite{pati2019inpainting}). Left: a one-bar melody. Right: uneven quantization bins. Bottom: representation sequence of this one-bar melody. The two strings in blue are the tonic and mode respectively.}
  \label{fig:representation}
\end{figure}

We convert the sheet music into note token sequences using the above scheme. Here we introduce two additional tokens, the tonic and mode, to represent the key signature of sheet music. These two tokens are put in front of the note sequences. Example of our representation is shown in Fig.~\ref{fig:representation}.

\section{Experiments}\label{sec:experiment}
\subsection{Datasets and Metrics}
We test our proposed Melody Structure Transfer Net (MSTN) on two public datasets. The Session dataset~\cite{sturm2016music} and the Wikifonia dataset~\cite{wikifonia}. The Session dataset consists of monophonic folk melodies in the Scottish and Irish style taken from the Session website~\cite{sturm2016music}. The original Session dataset consists of more than 48K melody pieces with ABC notation format~\cite{walshaw2011abc}. In this paper, we take the same subset as \cite{pati2019inpainting} from the Session dataset. Only melodies with 4/4 time signature are considered, and the pieces which consist of notes less than sixteen note are dropped. This results in approximate 21K melodies. The Wikifonia~\cite{wikifonia} contains more than 6500 lead sheets in MusicXml format~\cite{good2001musicxml}. We also select a subset from these music pieces according to the above standards and finally there are 3500 sheets. The chords track are dropped and only melodies are kept. These two datasets are split to $90\%$ training and $10\%$ validation. The training pieces are augmented with the pitch transposition strategy. All pieces are encoded into the sequence representation as stated in Sec.~\ref{sec:representation}.

\begin{figure}[h]
  \centering
	\subfigure[Wikifonia dataset]{
  \includegraphics[width=1.0\linewidth]{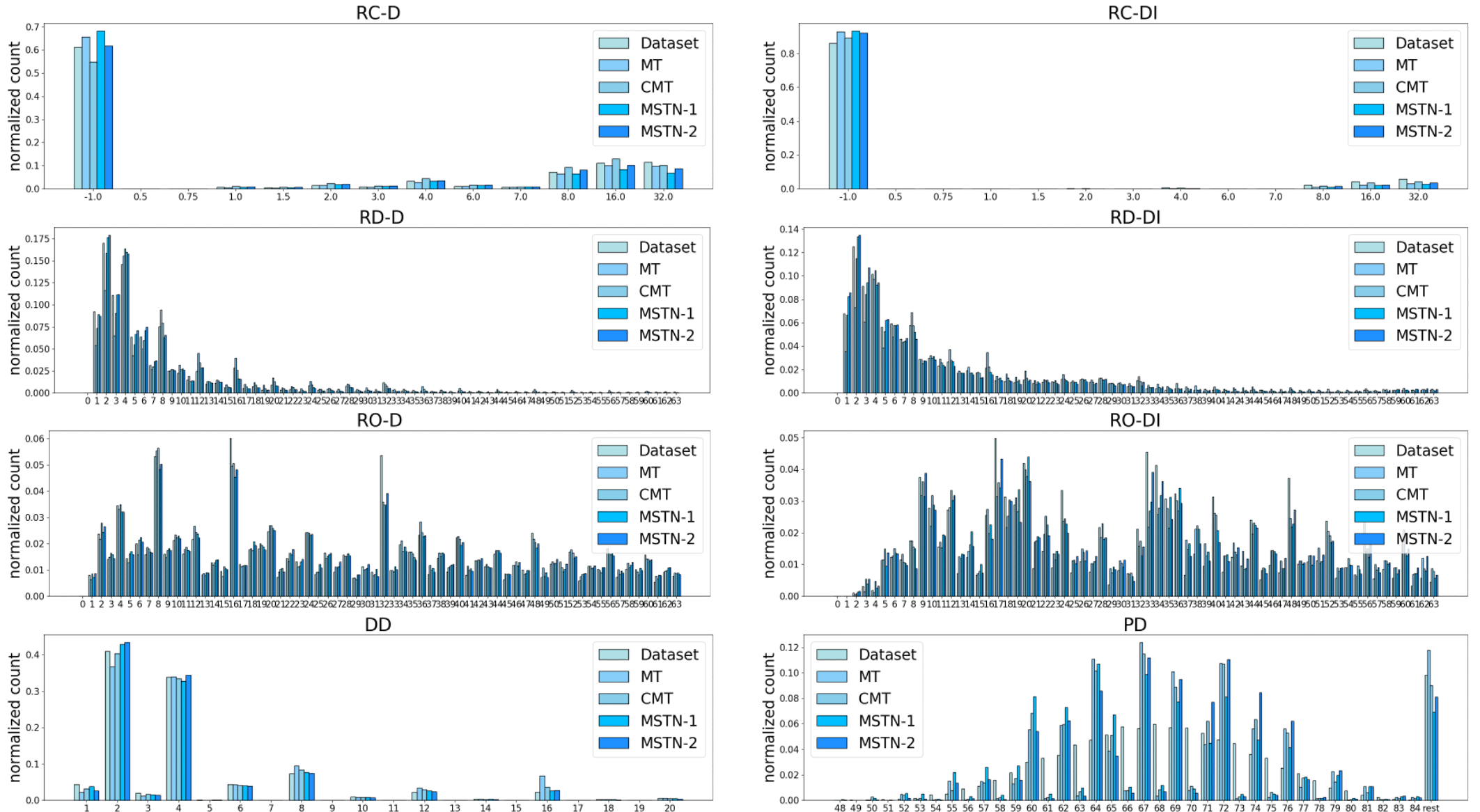} }
	\subfigure[Session dataset]{ \includegraphics[width=1.0\linewidth]{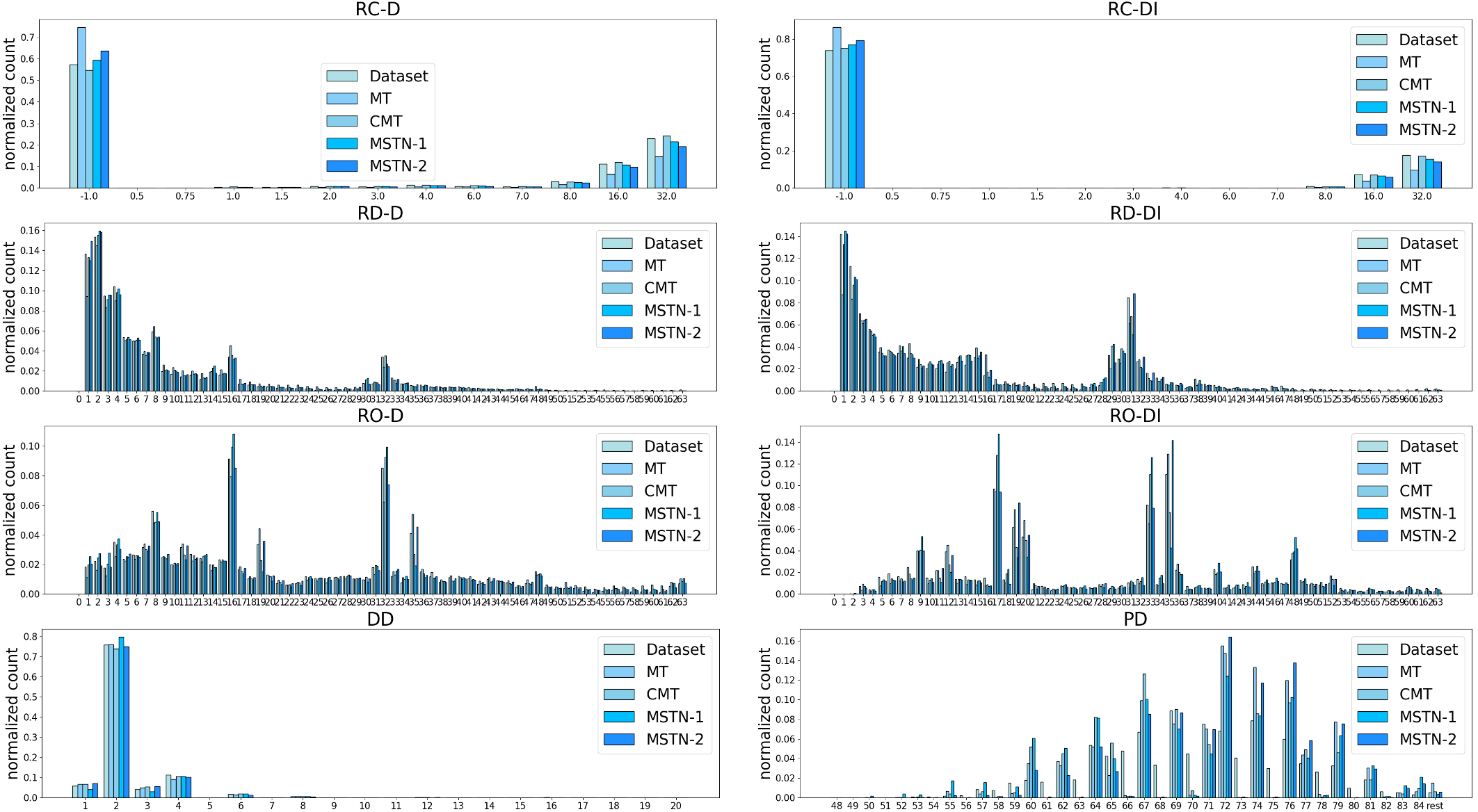}}
  \caption{Repeat-related statistics, pitch and duration distributions. }
  \label{fig:dist-wiki}
\end{figure}

To demonstrate that MSTNs can transfer the templates' structure to the generated ones, rather than simply remember the training templates, we adopt several structure related metrics~\cite{medeot2018structurenet}. We also
devise several metrics to evaluate the structure similarities and sample diversity. The metrics we used are as follows:

\textbf{1) Repeat Count (RC), Repeat Duration (RD) and Repeat Onset (RO):} Number of repeats corresponding to various lookback values, various durations, and various onset. These three kinds metrics are proposed in~\cite{medeot2018structurenet}, where only 16 bars of short music fragments are considered. We improve these measures to capture the repeat pattern across longer term by extending the lookback values (see Fig.~\ref{fig:dist-wiki}). These three metrics are calculated for the duration repeats (-D) and duration-interval repeats (-DI) separately, leading to 6 metrics -- see computing details in \cite{medeot2018structurenet}.
    
\textbf{2) Pitch Distribution (PD) and Duration Distribution (DD):} The statistics of various pitches and durations. These two metrics are popular in music generation tasks to compare the pitch and duration distribution similarity between the generated music sets and training sets~\cite{medeot2018structurenet}. 
    
\textbf{3) Rhythm Structure Similarity (RSS) and Interval Structure Similarity (ISS):} Here we adopt the self-similarity matrix (SSM), which is commonly used for the music structure analysis~\cite{jun2015music}. Despite that SSM is often calculated from the music audio, some researches~\cite{jhamtani2019modeling} applied it to the sheet music. In this paper we compute the SSM for rhythm and intervals separately at the bar level.
For each note in a bar, we can calculate its duration beat length and its interval from the previous note. We represent the rhythm pattern in a bar as list of (start time, duration) tuples, and represent the interval pattern in a bar as list of intervals:
\begin{equation}
\begin{array}{ll}
    Rhy(j) = [(st_1, d_1),(st_2, d_2),...,(st_m, d_m)]\\
    Intv(j) = [iv_1, iv_2, ..., iv_m]
\end{array}
\end{equation}
where $m$ is the number of notes in a bar, $st_{l}$ denotes the start beat position of the $l^{th}$ note in the bar, and $d_l$ denotes its duration. $iv_l = interval(iv_{l-1}, iv_{l})$ denotes the staff interval from the previous note to current note. If there is no previous note or the interval is inapplicable (eg. one of the note is 'rest'), the interval value is assigned as $null$.

With the rhythm and interval representation for each bar, we are able to calculate the rhythm and interval similarity between any two bars. And then the SSM for any given music piece can be obtained. We compute the rhythm similarity between two bars as the ratio between the duration of matched tuples and the total duration of a bar, and the interval similarity as the ratio of length of max matching string in two bars to the max length of these two bars:
\begin{equation}
 \small{
 \begin{array}{ll}
   SSM_{Rhy}(i,j) = \frac{\Sigma_{l\in ms}d_l}{Dur}, ms=\{l|Rhy(i)(l) \in Rhy(j)\}\\
   
   SSM_{Intv}(i, j) = \frac{max\ match\ length (Intv(i), Intv(j))}{max length (Intv(i),Intv(j))}
 \end{array}
}
\end{equation}
where $i,j$ are indexes of two bars, $Dur$ is the total duration of a bar, and for music pieces of 4/4 time signature, $Dur$ is 4 beat length. Then we can obtain a SSM of size $L*L$ for a music piece of $L$ bars. The SSM is a symmetric matrix as the similarity between bar $i$ and $j$ equals to is counterpart.

After computing the SSM for each music piece, we are able to evaluate the structure similarity between any two music pieces by compare their rhythm and interval SSM. The Rhythm and Interval Structure Similarity are calculated as the \textbf{root mean square error} between their corresponding SSMs.

\textbf{4) Rhythm Duplicate Rate (RDR) and Interval Duplicate Rate (IDR):}
It is necessary to check to what extent the model could generate new samples, rather than simply remember the training samples. It is also important to evaluate the diversities between the samples generated from a given template structure. Here we propose the rhythm duplicate rate and the interval duplicate rate to evaluate the diversity between the template and its generated samples ($RDR_{TAB}$ and $IDR_{TAB}$), as well as the diversity between the samples generated from the same template ($RDR_{AB}$ and $IDR_{AB}$). These duplicate rates are computed as the ratio of the number of bars with same rhythm or intervals between two samples to the number of bars in the samples.
\begin{table*}[tb!]
\centering
\resizebox{1\textwidth}{!}{
\begin{tabular}{|l|c|c|c|c|c|c|c|c||c|c|c|c|c|c|c|c|}
\hline
Datasets& \multicolumn{8}{c||}{WikiFonia}& \multicolumn{8}{c|}{The Session}\\
\hline
Statistics &RC-D & RD-D & RO-D & RC-DI & RD-DI & RO-DI &PD & DD &RC-D & RD-D & RO-D & RC-DI & RD-DI & RO-DI &PD & DD  \\
\hline
\hline
MT~\cite{huang2018music} & \textbf{0.004} & 0.091 & 0.014 & 0.029 &0.077 & 0.064 & 0.790 & 0.039& 0.073 & 0.040 & 0.038 & 0.054 &0.076 & 0.048 & 0.916 & 0.006\\
\hline
CMT~\cite{keskar2019ctrl} & 0.015 & 0.017 & 0.012 &\textbf{0.006} & \textbf{0.009} &0.053&0.569&0.009& 0.008 & 0.011 & \textbf{0.009} & 0.009 &\textbf{0.026} &\textbf{0.028}&0.943&0.006\\
\hline
MSTN-C (ours) &0.022 & \textbf{0.015}&0.015&0.033&0.013& 0.054& \textbf{0.481} &\textbf{0.003} &0.002 & \textbf{0.009}&0.015&\textbf{0.001}&0.031& 0.044& 0.880 &\textbf{0.003}\\
\hline
MSTN-U (ours) & 0.008 &0.016 &\textbf{0.010} & 0.021 & 0.014 & \textbf{0.048} & 0.707 & 0.009& \textbf{0.001} &0.010 &0.034 & 0.003 & 0.035 & 0.112 & \textbf{0.738} & 0.008\\
\hline
\end{tabular}
}
\vspace{-10pt}
\caption{KL-divergences between the training data and the generated melodies on the Wikifonia and Session dataset.}
\label{tb:kld-wiki-sess}
\end{table*}

\begin{table*}[tb!]
\resizebox{1\textwidth}{!}{
\begin{tabular}{|l|c|c|c||c|c|c||c|c|c||c|c|c|}
\hline
Datasets&\multicolumn{3}{c||}{WikiFonia} & \multicolumn{3}{c||}{The Session}&\multicolumn{3}{c||}{WikiFonia} & \multicolumn{3}{c|}{The Session}\\
\hline
\multirow{2}{*}{Models}&\multicolumn{6}{|c||}{Free Composition Mode}&\multicolumn{6}{|c|}{Continuation Mode}\\
\cline{2-13}
& MSTN-C & MSTN-U & CMT & MSTN-C & MSTN-U & CMT& MSTN-C & MSTN-U & CMT & MSTN-C & MSTN-U & CMT\\
\hline
\hline
\textbf{RSS}  &0.395&\textbf{0.335}&0.549&\textbf{0.422} &0.491&0.569&\textbf{0.441} & 0.442 & 0.558 & 0.551 &\textbf{0.537} &0.580\\
\hline
\textbf{ISS}  &0.371&\textbf{0.309}&0.482&\textbf{0.203}&0.229&0.313& \textbf{0.374} & 0.375 & 0.472 & \textbf{0.245} & 0.255 &0.322\\
\hline
\hline
$RDR_{TAB}$ &0.295&0.241&0.050&0.272&0.194&0.141 & 0.238 & 0.242 & 0.070 & 0.005 & 0.008 &0.013\\
\hline
$IDR_{TAB}$ &0.065&0.040&0.005&0.013&0.001&0.001 & 0.045 & 0.032 & 0.005 & 0.000 & 0.000 & 0.000\\
\hline
$RDR_{AB}$ &0.263&0.223&0.224&0.376&0.240&0.282& 0.425 & 0.444 & 0.245 &0.092&0.075 &0.060\\
\hline
$IDR_{AB}$ &0.040&0.030&0.060&0.004&0.005&0.007& 0.109 & 0.094& 0.065 &0.039&0.039&0.041\\
\hline
\end{tabular}
}
\caption{Structure Similarities and Duplicate Rates.}
\label{tb:smm}
\end{table*}

\subsection{Evaluation Protocol}
We evaluate the two versions of MSTN (i.e. MSTN-C and MSTN-U, corresponding to Fig.~\ref{fig:newatt3}(b) and (c), respectively. We also evaluate the implemented conditional music transformer (CMT). A baseline music transformer (MT) is also implemented and evaluated on several applicable metrics for clarity. 
All these models consist of 7 layers of self-attention blocks with 8 heads and 256 hidden states. The learning rate is $2e-5$ with 5 epochs of warm up. The max length is set to 100 bars (2400 time steps). The models are trained for 100 epochs on Wikifonia and 50 epochs on Session dataset. The parameter $\lambda$ in Eq.~\ref{eq: newatt1} and Eq.~\ref{eq: newatt2} is chosen as $0.1$ after we coarsely browse several values ranging from $0.001$ to $1$.

For evaluation, we generate two samples for each given template, which enables to compute the duplicate rates between samples generated from the same templates. The structure similarities between samples and their templates are calculated separately on the two samples and then averaged. Each of the music pieces in the training set has been taken as template. We generate samples in two modes: 1) free composition mode: trained models generate samples freely, without any given prime; 2) continuation mode: trained models generate subsequent sequences from a given one-bar motif. The metrics are computed on each set of generated samples, and then averaged on the whole dataset.

\begin{figure}[tbh]
  \centering
  \includegraphics[width=1\linewidth]{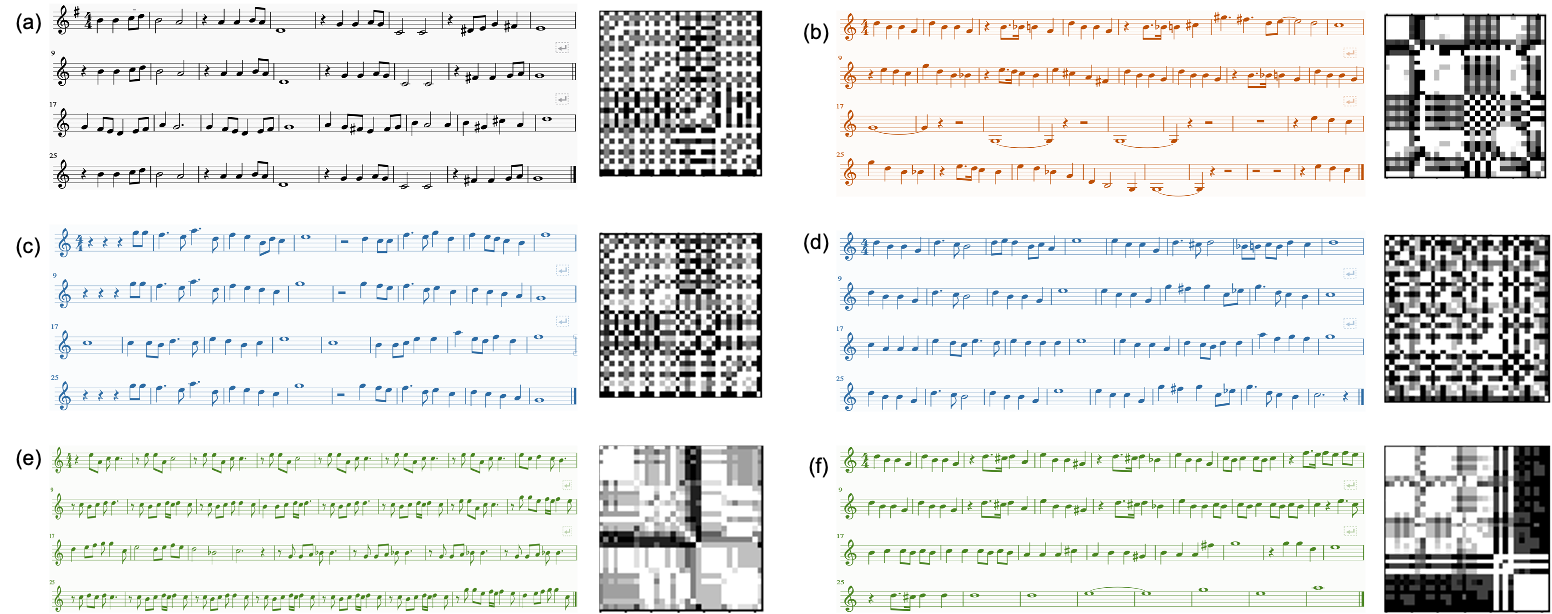}
  \caption{Melodies and their SSMs of (a) template melody, (b) melody generated with baseline music transformer in continuation mode,  (c) using MSTN-U in free composition mode, (d) using MSTN-U in continuation mode, (e) with CMT in free composition mode, (f) with CMT in continuation mode.}
  \label{fig:expscore}
\end{figure}
\subsection{Results and Discussion}
The statistics of the repeat-related metrics, as well as the pitch and duration distribution are given in Fig.~\ref{fig:dist-wiki}. The KL-Divergences on these distributions between the dataset and the generated set from each model are shown in Table~\ref{tb:kld-wiki-sess}. The calculation of these KL-divergences is the same as that in \cite{medeot2018structurenet}. To better understand the performance, we also present the evaluation statistics on the baseline music transformer model (MT). Table~\ref{tb:kld-wiki-sess} shows that the samples generated by CMT and the proposed MSTNs are closer to the dataset than those of the baseline MT. It is reasonable because CMT and MSTNs have more constraints than the baseline MT during the training process. These statistics do not differ much between the samples generated by CMT and MSTNs.

Table~\ref{tb:smm} shows the statistics (the lower the better) on structure similarities (RSS, ISS) and duplication rates (RDR, IDR). The samples of MSTNs and CMT generated in the free composition mode and the continuation mode are evaluated. In both free composition mode and continuation mode, MSTN-C and MSTN-U could generate samples which have more similar rhythm structures and interval structures to the template melodies than the CMT. This means that the proposed MSTNs can transfer the structure from the template melodies to the generated pieces well, while the CMT fails.

As the CMT samples has weak relationship with templates, their rhythm and interval duplication rates with the templates are also lower than MSTNs samples. But as to the duplication rates between samples, the CMT samples in free composition mode dose not perform better than MSTNs, which means that in free composition mode, the samples' diversities of proposed models does at least as well as CMTs. In continuation mode, it is also reasonable that MSTNs' samples are less diverse than CMTs' because the structure constraints of MSTNs are more tense than CMTs. Given the same motif, MSTNs will try to develop the motif according to the composition techniques of the template melody, which will definitely lead to similar generated samples. As for CMT, the model does not learn the structure embedding well, so it will not develop the motif according to the template composition techniques, and thus lead to diverse samples. So the lower duplication rates of CMT samples in the continuation mode also show that the CMT model can not transfer the structure well. Fig.~\ref{fig:expscore} show the melodies and theirs SSMs. The SSMs in this figure are the addition of the rhythm SSM and the interval SSM. Figure~\ref{fig:expscore}(a) shows a template melody. (b) and (d) are melodies generated using MSTN-U and CMT in free composition mode, respectively. (c) and (e) are melodies generated using these two model in continuation mode, where the first bar is the given motif. It is obvious that the SSM of MSTN-U melodies are much similar with the template's SSM, and SSM of CMT melodies are much different with that of template. We can see from (c) that the given motif is developed in to melody with a composition technique which is much similar to the template, where the bar 9-16 is similar to the bar 1-8, bar 21-24 is similar to bar 17-20, and bar 25-32 is similar to bar 9-16.

These results verify that the proposed MSTNs can transfer the structure of template melodies to generate new samples. The generated samples are of good diversity in the free composition mode, and their diversity will drop if the samples are generated from the same prime. MSTNs perform much better than the CMT. We upload 3000 sets of generated results in music xml format for the Wikifonia dataset as supplement materials. These pieces are generated with MSTN-U model in continuation mode, where the first bar is the given motif. Each folder is named with the corresponding template's name in the Wikifonia dataset, and for each template two samples are generated.


\section{Conclusion}\label{sec:conclusion}
We have proposed to transfer the structure of training samples for new music generation by a tailored self-attention mechanism, for long music generation. We also devise four quantitative metrics according to music theory. These four new metrics combined with eight existing metrics are used to evaluate our melody structure transfer model, with promising results. We will go deep and explore the structure transfer methods for polyphonic music generation in the future.
%



{
\small
\bibliographystyle{plain}
\bibliography{arxiv_2021.bbl}



}

\end{document}